\documentclass[aps,pra,preprint,groupedaddress]{revtex4}

\begin{document}
\title{On the Nature of Quantum Phenomena}

\author{Xiaolei Zhang}
\email{xzhang5@gmu.edu}
\affiliation{Deparment of Physics and Astronomy, George Mason University, Fairfax, VA 22030, USA}
\date{\today}

\begin{abstract}
It is shown that a coherent understanding of all quantized phenomena, including those governed
by unitary evolution equations as well as those related to irreversible quantum measurements, can be
achieved in a scenario of successive nonequilibrium phase transitions, with the lowest hierarchy 
of these phase transitions occurring in a ``resonant cavity''
formed by the entire matter and energy content of the universe. In this formalism, the physical
laws themselves are resonantly-selected and ordered in the universe cavity in a hierarchical manner,
and the values of fundamental constants are determined through a Generalized Mach's Principle.
The existence of a preferred reference frame in this scenario is shown to be consistent with the 
relational nature of the origin of physical laws.  Covariant unitary evolution is shown to connect
smoothly with the reduction of wavefunction in the preferred frame during quantum measurement.  The 
superluminal nature of quantum processes in the lowest hierarchy coexists with the 
universal speed limit obeyed by processes in higher hierarchies.  A natural quantum-to-classical
transition is also obtained which is stable against the diffusive tendency of the unitary quantum
evolution processes. In this formalism a realistic quasi-classical ontology is established for the 
foundations of quantum mechanics.
\end{abstract}

\pacs{03.65.Ta,03.65.Ca,03.65.Yz,03.65.Ud}
\maketitle

\section{Introduction}

Since the advent of Einstein's special relativity (SR) theory more than a century ago\cite{E1}, we have
become accustomed to the notion that fundamental physical laws do not
require a preferred or absolute reference frame.  The covariant transform relations obeyed by the relevant
physical quantities guarantee that these laws will take identical forms for all inertial observers in 
uniform motion with respect to one another.

With the dominance of this relational perspective, we tend to forget that the alternative view of a 
preferred reference frame is {\em also} consistent with the covariant nature of physical laws, 
and thus with all experimental results known to date, as long as the transformation relations between 
the preferred frame and other inertial frames remain that of the Lorentz transform relations.  
For example, the speed of light in vacuum $c$ could be first defined with respect to the
preferred frame, then other inertial observers simply measure and obtain the same speed of light because
of the Lorentz transform invariance implicitly obeyed by Maxwell's equations.
The merit of this latter point of view, however, lies in that certain classes of quantum
phenomena, i.e. those relating to the reduction of quantum mechanical wavefunction during
the measurement processes, appear to {\em require} a preferred frame to avoid uncertainty in the
triggering sequence of multiple widely-separated detectors as perceived by 
different inertial observers in relative uniform motion with respect to one another, and
to avoid the causal loops thus result\cite{DT1,DT2,B1,Eb1}.
The superluminal nature of the global wavefunction reduction during a quantum measurement\cite{Bell1,Bell2}
also appears to require the violation of general covariance which incorporates the speed of light 
limit for signal propagation.

In recent decades progress on both the experimental and theoretical fronts
brought into ever sharper focus the conflict between quantum phenomena as we 
know them, and the current state of our understanding 
according to the orthodox explanations of quantum mechanics
regarding fundamental issues.
On the experimental front, experiments such as the independent laser source interference experiments 
carried out by Pfleegor and Mandel\cite{PM1, PM2} hinted at the existence of a coherent quantum radiation
field ``out there'' in the environment, independent of which source(s) had emitted the quantum radiation, a view
which is also supported by the well-known relation between the coherent length/coherent time of an 
electromagnetic field with its bandwidth irrespective of the source of the radiation\cite{MW}; 
whereas experiments confirming the delayed-choice nature\cite{J,WL1} of quantum measurement process 
raised issues on the validity of temporal causality\cite{WL2,TF,SB} if the traditional Copenhagen
interpretation is to be retained; furthermore, results of the experimental tests of Bell's 
inequality\cite{Bell1, Bell2} support the reality of faster-than-light signal propagation
in wavefunction-reduction type of quantum processes, and also bring into question which observer's frame such
a reduction is happening with respect to; phenomena such as the Aharonov-Bohm effect\cite{AB}, 
Aharonov-Casher effect\cite{AC}, as well as Josephson effect in superconductivity\cite{JS}
further reinforced the impression that the phase field in the quantum mechanical wavefunction has a 
realistic existence, and is globally defined with repsect to a universal frame.  On the 
theoretical front, several works from different perspectives\cite{DT1,DT2,B1,Eb1} showed that the 
internal structure of quantum mechanics and the totality of quantum phenomena are
inconsistent with a thoroughly Lorentz invariant formulation, and a preferred reference frame 
is required if we were to arrive at a self-consistent description of quantum phenomena.
This requirement, however, is apparently in conflict with the successful covariant formulation
of much of the modern physical theories, as well as their experimental validation.
A complete and self-consistent theory of quantum processes must therefore be able to reconcile
the exisitence of the preferred frame on the one hand, and the covariance nature of
physical laws on the other.  Such a theory must also be able to reconcile the superliminal
signal propagation during quantum wavefunction reduction with the finite propagation speed of
most of the physical processes, in both the quantum and the classical world.

There is thus the need to carefully reexamine the fundamental issues in both quantum
theory as well as other branches of physics, in order to arrive at a coherent and
consistent understanding of the physical phenomena they describe. 
The current paper, together with other preliminary writings by the present author\cite{XZ1,XZ2}, 
describes the initial results of a systematic exploration  
which aims to allow the seemingly peculier quantum phenomemna, in particular the nature
of quantum measurement processes, to be explained in a realist and intuitive fashion, 
and allow a coherent framework to be erected linking quantum and classical physics. 

\section{A New Ontological View on the Nature of Quantum Phenomena}

One of the conceptual influences Einstein received in developing
his theory of general relativity (GR)\cite{AE1} was Ernest Mach's conjecture that mass and
inertia of a given body resulted from the influence of all the matter content 
in the universe\cite{Mach} (see, e.g., p.~546ff of \cite{MTW}).  In the final mathematical
formulation of GR, however, the relation to this so-called
Mach's Principle is reflected only in the implementation of the
equivalence principle which allowed the global mass distribution to affect
the local spacetime curvature, and allowed the curvature in turn to affect the trajectory 
of a local test particle.  There is no explicit prescription of how to
generate local mass and inertia from the cummulative influence of other mass (both nearby and far-away)
in the universe.  Therefore GR is only partially Machian in terms of Mach's
original conception, and the effect of the neighboring
mass distribution on the quantitative value of inertia does not seem to be borne out
either from the theory or from experiments\cite{Wb}.  Thus the detailed fashions
that the spirit of Mach's principle are to be implemented have yet to be established.

GR and its relation to at least the gross features of Mach's principle
point to the possibility that other fundamental laws of nature may be relational
as well, and their precise forms might have resulted from the mutual interaction
of the matter and energy content in the universe.  The existence of quantized phenomena
described by univeral constants such as Planck's constant, and fundamental particle
properties such as electron's charge and mass which are universal, further motivates
the idea that the global and instantaneous interactions of
all the matter and energy content in the universe may have been the underlying cause
for a resonant generation of quantized properties of fundamental particles
as well as quantum and classical laws.

The universe is an open and evolving system.  In classical physics such a system
admits nonequilibrium phase transitions which result in the formation of
so-called ``dissipative structures''\cite{Prigogine1977, Zhang1998}.
If we insist on a realist interpretation of the quantum processes, and
on a quasi-classical ontology for quantum phenomena which connects with
intuition, we are naturally led to seek the origin of the universal quantized behavior
in the totality of the universe we inhabit which is a self-contained (and in this
scenario necessarily finite), open, and nonequilibrium system, in perpetual evolution from
the past to the future.  These considerations, as well
as others, formed the motivation for the new ontological view of quantum processes,
the main ideas of which are summarized as follows:
\begin{itemize}
\item It is assumed that a generalized ``Mach's Principle'' governs the operation 
of the physical universe. Features of physical interactions, including the values 
of fundamental constants and the forms of physical laws, are determined by the 
global distribution of all the matter and energy content in the universe and by
their mutual interactions.
Such a view is shared by many contemporary working physicists\cite{Barbour1995, Sachs2003}.
Consistent with this view is the existence of a single universal reference frame which
is established by the matter and energy content of the universe,
yet the laws of physics in such a universe can remain relational/covariant
because this absolute frame is defined by the entities in it collectively.
\item The quantized nature of fundamental processes originates
from nonequilibrium phase transitions in the (large but finite) universe
resonant cavity.  The substratum of the quantum phase transitions is however itself
continuous and thus has an infinite number of degrees of
freedom.  The usual quantization procedure by enforcing
the canonical commutation relation is equivalent to establishing modal
closure relation in the universe resonant cavity.  Uncertainty relations
are the phenomenological derivatives of the corresponding commutation
relations and thus have no independent fundamental significance.
\item Due to the successive nature of phase transitions
the physical universe is organized into hierarchies in
both its laws and its phenomena.  The hierarchy in the
physical laws is reflected in the spontaneous broken 
gauge symmetry to obtain physical laws in the different energy scales, as
well as in the division of the domains of validity
of laws for governing different branches of physics and chemistry.
The precise fashion of the operation of laws in the lowest hierarchy 
is mostly unknowable to human beings, since the lowest hierarchy processes 
communicate influences in superluminal fashion and can instantaneously sense
the global distribution of the entire matter content
in the universe.  These lowest hierarchy laws and processes are responsible for producing
the fundamental constants, the univeral forms of first-hierarchy laws
and their transformation relations, as well as for the global nature
of the quantum mechanical wavefunction collapse. 
The hierarchical ordering in physical phenomena, which is a direct
result of hierarchical ordering of laws, is responsible for forming
the quantum and classical division, as well as the micro- and macroscopic
division.  Macroscopic structures, being stable resonance
features in the large scale, are capable of resisting the
diffusion/smearing tendency of pure quantum states governed by
Schrodinger-type wave equations.
\item A quantum mechanical wavefunction describes the substantial
distribution of the underlying matter of a specific modal type
in the configuration space (this includes the kinds of wavefunctions
which have simultaneous projections in the more abstract modal-basis directions
as represented by the Dirac-type rather than the Schrodinger-type wavefunctions).  
Its absolute square gives the probability
for obtaining a particular result in the measurement phase transition.
Its phase encodes the influence of the environment including the
different types of force fields, and the gradient of the phase determines the
subsequent evolution of the wavefunction.  The probabilistic element
is thus removed from the ontology of quantum mechanics, its appraent influence
to the result of quantum measurement is due to the role played by the global
universe environment in the measurement phase transitions, and the universe
effectively has infinite degrees of freedom which prevents an exact prediction
of the measurement outcome.
\item A quantum measurement process happens under appropriate
boundary conditions such that an irreversible nonequilibrium
phase transition is induced in the joint system of the object being
measured, the measuring instrument and the rest of the universe.
Every irreversible quantum measurement process is effectively also the creation
process of new quantum mechanical particles/resonances, rather than
the capture process of the incoming quantum mechanical particles.
What is being propagated by the incoming quantum mechanical wavefunction
into the measuring instrument is only an influence field to the
total collapse, which together with the boundary conditions set up by
the measuring instrument and the rest of the universe determines the
outcome of the collapse, or new resonant mode formation. 
Yet this influence field is substantial, rather than
ficticious as in the de Broglie-Bohm pilot wave theory\cite{dB,BM}.
The causal connection of a quantum measurement is globally distributed, rather
than linear and sequential, though temporal causality remains strictly true in the
preferred reference frame with respect to which the wavefunction is anchored 
(but not necessarily at rest) and the wavefunction collapse happens.  
The collapse of the wavefunction during quantum measurement 
happens globally and superluminally.
\item In this picture the vacuum fluctuations are the ``residuals''
after forming the ``whole'' numbers of quasi-stationary modes in the open,
nonequilibrium universe cavity.  The fluctuations in the vacuum reflect
the continuous energy exchange with the ``whole'' modes needed to
sustain these whole modes, likely a quantum version of the well-known
``fluctuation-dissipation'' theorem satisfied by all the non-equilibrium 
quasi-steady-state structures in the classical regime.
\end{itemize}

One of the important corrolaries of the above set of hypotheses is that
the time axis of the four-dimensional spacetime continuum has now a special
status in the preferred frame compared to the three spatial axes, because 
quantization (or the formation of a new quantum resonance) happens in the 
spatial domain at a certain universal time.  This and other important
implications of the new ontology on the nature of
quantum as well as classical processes will be discussed
in subsequent sections.

\section{Implications on the Foundations of Quantum Mechanics}

It is well known that quantum interaction is not a
fundamental force of nature (which includes gravitational, electromagnetic, weak and
strong forces). It rather appears as a set of meta-laws which 
govern the collective behavior of the microscopic substratum under the influence 
of interacting forces to form collective quantum resonances, and to evolve them.  
The manner in which we treat quantum interactions and transitions is in essence a modal approach, 
and we use our knowledge of the collective properties of the input states and the 
Hamiltonian, together with appropriate boundary conditions, to predict the subsequent evolution of the states.  
In a way, in this kind of calculations we have already incorporated our insights and prejudices
regarding the physical processes we are trying to model.
Therefore in addressing foundation issues we must be careful to separate the mathematical
formalism which contains idealizations and approximations, from the underlying physical processes
which often contain more subtle complexities.

\subsection{Meaning of the Quantum Mechanical Wavefunction}

The quantum mechanical wavefunction or the so-called probability wave is a 
realistic physical entity in the new interpretation.
The fact that the wavefunction usually exists in the 3n-dimensional
configuration space does not pose a problem to the substantiality
of the wavefunction, since this signifies only
the interchange and interrelation of the parts and parcels of the
modal content among a multi-particle quantum state.
The dispersion of the position-eigenstate wavepacket\cite{Schrodinger1926}
is here seen as a natural tendency for a localized ``particle''
to evolve towards a plane-wave-like momentum eigenstate.

Whereas the amplitude of the wavefunction is connected to the modal
density in the configuration space, and thus leads to the
probability of a particular measurement outcome,
the spatial variation of the phase of the wavefunction characterizes
the probability flux, and thus determines its subsequent evolution\cite{Sakurai1985}.
The phase of the quantum mechanical wavefunction also carries imprints of
various kinds of interaction fields, and is the medium through which
force fields apply their influence to the wavefunction.

Both the amplitude and the phase of the
quantum mechanical wavefunction acquire substantial meaning in our picture,
and the probabilistic element is removed from the foundations of quantum
mechanics.  Its apparent presence in the quantum measurement process is viewed
as a result of the sensitive dependence on initial/boundary conditions of the
non-equilibrium phase transitions in a many degrees-of-freedom system, which
is effectively the whole universe.

\subsection{Unitary Evolution of the Quantum States}

The wavefunction of a quantum observable usually spreads out
in infinite space, and the interaction between the different
quantum modes are global in nature. Therefore
quantum mechanics is formulated in the Hilbert space which
is a natural domain to describe global modal relations.
                                                                 
Traditionally, a general quantum evolution is considered to involve the two-stage
process of a continuous unitary evolution 
which is covariant and energy conserving, followed by a discontinuous
``wavefunction collapse'' which leads to the reduction of the
wavefunction into an eigenstate of the measurement observable.
It was the second step, i.e., the wavefunction collapse process,
which acquires a global and irreversible feature in the new ontology.
                                                                   
Even during the unitary evolution phase certain features of global
and irreversible interactions are subtly and implicitly contained.
For a state which is not an eigenstate of the energy operator,
expanding it in terms of the eigenstates of an observable $\bf{\alpha}$ that
commutes with the Hamiltonian H, we have\cite{Sakurai1985}
\begin{equation} 
\Psi(\vec{x}, t) = \int d^3x' K(\vec{x},t, \vec{x'},t_0) \Psi(\vec{x'},t_0)
,
\label{eq:1}
\end{equation}
where
\begin{equation}
K(\vec{x},t; \vec{x'},t_0) = \sum_{{\bf \alpha}} <\vec{x}|{\bf \alpha}>
<{\bf \alpha}|\vec{x'}> \exp 
{{ {-i E_{{\bf \alpha}} (t - t_0)} \over \hbar}}
,
\label{eq:2}
\end{equation}
which represents the transition amplitude between the different energy eigenstates
as time evolves.  Therefore, as the
unitary evolution proceeds the relative weights of the constituent
energy eigenstates vary with time, i.e., there is a kind of
transition process happening continuously between the constituent eigenstates.
The Hamiltonian function $H$ is what drives the evolution (through its role in the
determination of the energy eigenstates and in changing the weights of the different
eigenstates with time), and the potential function
$V$ contained in the Hamiltonian encodes the global influences which are averaged
and idealized as the influencing potential.  The back-reaction
of the evolution of the wavefunction on whatever constitutes the sources of $V$ is
however not modeled.  These idealizing elements of the unitary formulation
conceal the true irreversible behavior in a subtle manner, and is to a large
extent responsible for the cause of 
paradoxial explanations of pheonomena such as interaction-free measurements\cite{IF1,IF2,IF3}
which we will address later.  In a way, something akins to ``one can never twice step
into the same stream'' happens here during the unitary evolution, in that the
universal environment (including the form of and the constants in the
fundamental laws) itself is irreversibly evolving as an open and nonequilibrum system,
true unitary evolution in the sense of perfect reversibility does not exist.

Even though an eigenvalue of a stationary
state is a constant (call it $\alpha$), the eigenfunction is in general
time dependent (the particular form of the evolution of the
eigenfunction, if it is a simultaneous eigenstate of the Hamiltonian, 
is $|{\bf \alpha}> \exp ({{-i E_{\alpha} t} \over {\hbar}}$)\cite{Sakurai1985}).
Therefore a quantum mechanical stationary state is a kind of dynamical
equilibrium in constant (quasi-periodic) evolution, including constant exchange with
the quantum vacuum, consistent with the nonequilibrium
stationary state picture we have proposed.

In practice, apart from stationary states and freely-evolving
wavefunctions, the density matrix formalism\cite{VonNeumann1927} 
had also been employed to describe quantum systems that are thought to be
statistical mixtures.  In the current ontology, no physical system
is actually in a statistical mixture state.  The apparent
success of the density matrix approach is understood as the intrinsic
harmonic nature of the evolution of the parts and parcels of constituent
quasi-stationary states -- thus the effect of time averaging
mimics the effect of ensemble averaging.

\subsection{Nature and Extent of Quantum Measurement}
                                                        
A quantum measurement is in general a non-local process. 
In our new ontology, quantum measurements involve not only the objects being measured and
the measuring apparatus, but also involve
the ``give and take'' with the rest of the universe. 
This ``give and take'' with the rest of the universe accounts for the apparent
violation of energy conservation in many quantum measurement 
processes\cite{XZ1}, and also underlies the probability element
and the projection/non-unitary nature of the quantum measurement.

A pure quantum mechanical resonance (such as a photon) remains a
modal resonance spread out in space (and often joins the reservoir of
the collective oscillation as represented by the classical fields) until 
the moment of detection.  The detected particle is in general no longer the same particle
as during propagation because of the exchange with the universal background,
and because in cases when the occupation number of the photon in the incoming field
is large a photon cannot be considered to have an independent existence during
propagation because it was part of the collective
oscillation field of the photon reservoir.

The strongest support for the involvement of the universe resonant
cavity during the quantum measurement process is actually the constancy
of emerging elementary-particles' characteristics in different types of
physical processes (such as the constancy of electron charge and mass
no matter the electron is created out of neutron decay, or else out of the
electron/positron pair production from energetic photons).
Without a global resonant cavity to determine
the modal characteristics, we would not have such universal entities as elementary
particles, or the fundamental constants themselves.
The identity of the properties of fundamental
particles is the result of their being the same global mode,
and the properties of these particles are created at the
moment of the measurement phase transition, since before this transition
an electron, say, does not already exist inside a neutron.
                                  
The nature of quantum measurement in our ontology is a global
phase transition which happens as a single and instantaneously process.
There is not a two-step ``decoherence'' followed by ``collapse by the
consciousness of the observer'' process as advocated in the von Neumann
measurement theory\cite{VonNeumann1932} and its derivatives.
The comparison of our current quantum measurement interpretation with other
existing proposals of quantum measurement can be found in \cite{XZ1, XZ2}.

\subsection{Wave-Particle Duality.
Photon and Classical Electromagnetic Field}

The wave-particle duality is manifest most clearly in the de Broglie
relation $ p = h/\lambda$, and the Planck relation $ E=h \nu$:
each equation on one side indicates a pure wave characteristic ($\nu$ and $\lambda$)
and on the other side a pure particle characteristic ($E$ and $p$).
The seemingly contradictory characteristics are easily clarified in the
new ontology:  a particle is more of a pure resonance
when it is in the form of a wave mode and is spread out.  When it is a localized particle,
it is in a mixed resonant state, or the superposition of pure states (signified
by a broad frequency spectrum).
                                                                  
In fact, in quantum field theories, only the fields are localized,
but field quanta are spatially extended.  These spatially distributed
field quanta arrive from the first approximation of the solution
of field equations in the non-interacting limit and
are the source of the name ``particle''\cite{Cao1998}.

The well known photon fluctuation formula obtained by Einstein\cite{Loudon}
contains two separate terms,
one contributed by the field amplitude fluctuation and one by the particle
number fluctuation.  This indicates that in the free propagating state the photon
number states/modes are indeed being excited, yet those modes are going through constant energy
exchanges among themselves as well as with the vacuum environment, 
i.e., the photons in those modes are
not well-separated entities but part of the collective wave oscillations.
                                                                  
\subsection{Uncertainty Principle and Commutation Relations}
                                                               
In general, the uncertainty relations can be shown to originate
from the corresponding equality relations
linking the commutators and anticommutators
of the quantum observables A and B as\cite{Sakurai1985}:
\begin{equation}
|< \Delta A \Delta B>|^2 = { 1 \over 4} |<[A, B]>|^2 +
{1 \over 4} |<\{\Delta A, \Delta B\}>|^2
,
\label{eq:11}
\end{equation}
where 
$[A, B]$, $\{\Delta A, \Delta B\}$
denote the commutator and anticommutators 
of the respective quantitites they enclose.
Furthermore, from Schwarz inequality
$ 
< \Delta A>^2 <\Delta B>^2 ~~ \ge
|< \Delta A \Delta B>|^2
,
$
the usual uncertainty relation can thus be arrived at:
\begin{equation}
< \Delta A^2> <\Delta B^2>
~~
\ge 
{ 1 \over 4} |<[A, B]>|^2
.
\end{equation}
Therefore, the well-known uncertainty relations are direct consequences of the 
non-zero-valued commutation relations, which are themselves deterministic.
The uncertainty principle itself seemed to later acquire more
prominence in the discussions of quantum phenomena mainly because of its
intimate relation to the probabilistic outcome of quantum
measurements.

\subsection{Quantum Vacuum}

After quantizing space with a set of modes using the commutation or
anticommutation relations, we expect to end up with some leftovers,
as is typical for the formation of nonequilibrium quasi-stationary modes\cite{Zhang1998}.  
We propose that the leftover fractional modal contents of
the quantization process are the constituents of vacuum fluctuations.
                                                                         
The vacuum field fluctuates because the resonant components keep evolving
in the nonequilibrium universe cavity, just as in another example of
such a nonequilibrium dissipative structure, that of the spiral structure in
galaxies\cite{Zhang1998}, where the individual star's trajectory keeps evolving
and moving in and out of the spiral density wave pattern even though total energy is
conserved among the wave mode and the stars, and the spiral density wave mode is meta-stable.
                                                             
Many effects which so far have been attributed to the quantum
vacuum, such as the Casimir force, the virtual particles and the lifetime of atomic levels,
as well as the spontaneous emission of radiation, can equally be thought
of as due to the influence of the rest of the matter distribution in the
universe.  For example, the addition of metal plates in the Casimir effect changes the boundary
condition of the entire vacuum, force is thus needed to put the plates in.
Virtual particles related to the quantum vacuum are those which appear in a quantum electrodynamic (QED)
calculation and do not satisfy energy and angular momentum conservation\cite{Sakurai1967}:
They are ``not on the mass shell'' and are represented by the internal lines
in Feynman diagrams.  Their existence is another indication
that a quantum phase transition involves the rest of the universe to
``close the loop'', and the conservation relation is restored for resonant
interactions only when the phase transition is complete (this last condition
is in fact not yet met by the current perturbative QED, in the strong
coupling case or in higher order calculations, likely indicating the
inherent inability for a local field theory to become self-consistent).
                                                                   
This view of the vacuum also provides a possible explanation of why
some of these effects (including Lamb shifts,
Casimir effects, spontaneous emission, van der Waals forces, and
the fundamental linewidth of a laser) can be explained equally successfully
by adopting either the vacuum-fluctuation point of view
or the source-field point of view\cite{Milonni1994}.
                                                                      
The relation of field quantization and vacuum fluctuation  may also be
related to the ``fluctuation-dissipation theorem''.  The dissipative
leak into the vacuum is needed to maintain the stability of the
nonequilibrium modes.   The substratum needs to be constantly
evolving in order for the fundamental resonances to be stable.
So the un-saturatedness and the constant evolving nature of the universe
maybe a prerequisite for setting up the fundamental laws and structures that we
observe today.
                                                            
\subsection{First and Higher Order Coherence of Photons and Atoms.
Identical Particles}

When Dirac commented that a photon interferes only with itself\cite{Dirac1958},
he referred to the first-order coherence property of the photons\cite{Loudon}.
Subsequent intensity interferometry experiments\cite{Brown1954}
had revealed that photons do interfere with one another,
which are the higher order coherence properties of photons.
Such first and higher order coherence properties were also observed
for atoms in the atom interferometry experiments\cite{Berman1997}.
                                                                   
In the current ontology, the first order coherence of the atoms and
photons reveals their underlying wave and modal nature, whereas the
higher order correlation is a manifestation of the finite-Q nature 
of the universe resonant cavity, which results in the ``non-pure'' spatial modes
with mutual entanglement.
Due to this entanglement (as reflected in the Bose-Einstein or
Fermi-Dirac statistics, for example), after the emission process
a photon has a tendency to merge back to the universal ``reservoir''
of the background photon flux during propagation, unless the photon
flux is so low that it can be described as spatially and temporally separated 
monophotonic states, in which case its degree of second order coherence 
$g^2(0) <1$ as is appropriate for photons in the non-classical photon number states\cite{Loudon}.
The analytical expressions for the degree of second-order coherence
for bosons and fermions show different expressions according to
their respective wavefunction symmetries\cite{Scully1997},
and these statistics are only meaningful when the particle
flux is high enough.

\subsection{Quantum-to-Classical Transition. Resolution of the ``Schrodinger's Cat Paradox''}
                                                             
The physical world described by the reversible and unitary quantum evolution equations is forever changing
and diffusing.  This image conflicts with the picture we have of classical objects, which are well-defined
in shape and stable in their properties.  If quantum theory underlies the explanation of
classical objects as well, this transition from the quantum to the classical behaviors will need to be explained.

Under the new ontology there is no longer a dichotomy between
the classical and the quantum world.  Classical systems consist
of subunits where ``wavefunction collapse'' has already been induced
by nature, through naturally occurring boundary conditions.
After a spontaneous phase transition, the overall system will be in a
quasi-stationary state and thus is stable, though its constituent parts
may still evolve in a harmonic fashion as shown before for the
energy eigenstate.  A macroscopic object in general does not possess
an overall quantum mechanical wavefunction that freely evolves as a single entity
because its internal structural stability requires the
various collapsed subcomponents to evolve as approximately separated and well delineated
subsystems under the average boundary conditions of their environments,
though in the truest sense the entire universal environment is connected
in the lowest hierarchy to form the fundamental constants and laws.  There is thus the subtle
interplay between isolation and connectedness of the different constituents
in the structures of macroscopic systems.
                                                                  
The spontaneous nature of phase transitions in natural systems
helps to resolve the ``Schrodinger's Cat'' type of paradoxes\cite{Schrodinger1935} since a
naturally occurring ``quantum measurement'' does not have to involve a conscious observer
to ``collapse the wavefunction'' at the moment of his observation.
The cat in question was already in a definitive Live
or Dead state before the observer opened the box, and not in a 
linear superposition state of the kind: $a \cdot {\rm Live} + b \cdot
{\rm Dead}$.  

The phase transition view also explains the stability
and reproducibility of natural orders, i.e., the results
of non-equilibrium phase transitions are insensitive to the
{\em details} of the initial-boundary conditions, and depend
only on the gross nature of these conditions\cite{Prigogine1977, Zhang1998}.

\section{The Hierarchical Ordering of Quantum and Classical Mechanics}

We now turn to look at the implications of the new ontology on the structural
organizations of physical laws.  There are three basic ingredients
for the new organizational principles:
\begin{enumerate}
\item The physical interactions in the universe are organized in a hierarchical
fashion, with each hierarchy governed by its own set of laws and transformation relations.
\item In the most fundamental hierarchy (which we shall call the zeroth hierarchy)
there is effectively instantaneous propagation of the influences
from the global distribution of matter and energy, which generate first order physical
laws as we observe them, as well as fundamental constants.  The wave function collapse
phase transition is also a zeroth-hierarchy phenomenon.  The detailed
operation of zeroth hierarchy processes in generating first hierarchy
laws are difficult to perceive by human observers because of the instantaneous nature
of the interaction and change.
\item There is the coexistence of an absolute reference frame 
defined by the entire matter and energy content of the universe, together with the relational nature
of interactions among the same matter and energy content.  This is made possible 
in part by the equivalence of the Lorentz and Galilean transformations when the 
signal propagation speed is infinite, so the relational and the absolute could coexist (Appendix A).
The appearance of the frame-independent (or covariant) nature of the first
hierachy laws is perfectly consistent with the preferred-frame nature of the
zeroth hierarchy operations because the first hierarchy processes are emergent
phenomena from the zeroth hierarchy, and in both hierarchies the laws are relational, and
only the maximum signal propagation speed is different in these two hierarchies.
\end{enumerate}
We now comment briefly on certain elements in the above basic ingredients in the
organization of physical laws.

Why would nature need ``instant messaging'' to set up the zeroth hierarchy laws? 
This is because the universe is enormous, even though finite as demanded by this theory
(so that the fundamental constants themselves would come out finite through resonance selection),
and it takes too long for the round-trips needed to achieve resonant selection if we have only a 
finite propagation speed in the zeroth hierarchy, and in that case the resulting first-hierarchy 
laws won't be universal in form, since they would be responding to local gradients of 
interactions.  Since the universe we reside in is large but finite,
it is conceivable that the zeroth hierarchy signal propagation speed is only
effectively infinite, or larger than anything we can conceive, yet is not truly
mathematically infinite. In the following discussion, however, we will for the most part
not distinguish this subtle difference between the effective and the true instant-communication
in the zeroth hierarchy, except in discussions dealing with long term 
evolution of the universe and physical laws. 
Similarly, the variational approach for the selection of physical laws
only works if all possible paths are explored simultaneously/instantaneously.
This ``instant messaging'' nature is also the underlying
reason for the validity of many global statistical laws such
as Einstein's statistical treatment of spontaneous and stimulated emissions,
and the quantum statistics of identical particles.  These laws would not have had
their universal validity if the relevant ensemble is not sampled instantaneously.

Why then do the first-hierarchy laws need to be Lorentz invariant? Because laws must be
neutral to all observers in a homogeneous and isotropic universe whose
laws are set up by relational interactions.  If the speed of maximum signal propagation
is finite for the first-hierarchy (as it should be, for otherwise we will not be
able to observe changes and evolution, and time would lose its meaning)
and is also universal, that naturally lead to the Lorentz
type of transformation relations for the first-hierarchy laws\cite{E1}.

In this scenario in the zeroth hierarchy spacetime is flat, continuous
and absolute, and is decoupled from the dynamics it generates.
The infinite signal propagation speed in this hierarchy enables universal
time synchronization across space (Appendix A).
With the hierachies thus set up, all physical laws on the macroscopic scale, 
including SR and GR, become emergent properties in the universe cavity.
The Lorentz transform satisfied by the first hierarchy laws
mixes the space and time axes due to the finite signal propagation speed
in the first hierarchy, and a large part of this mixing
can be regarded as a perceptual effect, yet it
has real consequences since this perceived effect is what contributes to the mean-field
type of potentials which inflence quantum mechanical state evolution and
which in a relational manner allows the physical laws in the first hierarchy
to be transformed in a covariant manner across the entire spacetime.
From these first hierarchy laws (which are univeral and governed by covariant
transformation relations) there can emerge second-hierarchy laws such as
those we observe in  solid state physics\cite{LP}.  These second-hierarchy laws
are based on zeroth and first hierarchy laws, but their generation is
further influenced by local boundary conditions.

The finite propagation speed existing in the first hierarchy laws can be generated 
from the infinite propagation speed in the zeroth hierarchy. An analogous example of
this nature in our familiar physical world is the generation of the
spiral density wave modes in disk galaxies\cite{Zhang1998}, where the gravitational 
influence is modeled as propagating instantaneously (it uses pure Newtonian
gravitational physics), yet the group velocities of the branch
wave trains are found to be finite, and are determined by the
so-called ``basic-state'' properties, which form the effective boundary
conditions of the galactic resonant cavity that these density wave modes
emerge from.

The unpredictability element in quantum mechanical measurement events 
whose probability is nontheless predictable is likely to be partly a result of the interplay 
between the Lorentz-invariant mean-field laws (first hierarchy), which evolve
deterministically and form the boundary-condition potentials, and the
absolute-frame wavefunction collapse (zeroth hierarchy), which are simultaneously
influenced by all the infinite degrees-of-freedom processes distributed throughout
the universe.

\subsection{Origin of Physical Laws}

The idea that matter throughout the universe collectively determines
the inertial properties of local matter was the original Mach's principle\cite{Mach}.
A generalized version of Mach's principle demands that all physical
interactions be relational.  This characteristic is reflected in
Feynman's formulation of a significant
fraction of dynamical laws (both quantum and classical)
as an integral of the classical action $L_{classical}$
over all possible spatial paths and histories ${\cal D} [x(t)]$ \cite{Feynman1965} i.e.,
\begin{equation}
<x_N,t_N|x_1, t_1> = \int_{x_1}^{t_N}
{\cal D} [x(t)] \exp [ i \int_{t_1}^{t_N} dt  {{L_{classical} (x, \dot{x})}
\over {\hbar}}]
.
\label{eq:fm}
\end{equation}
The above space-time formulation explicitly demonstrated
that the quantum mechanical amplitude at any given location and time
is the sum of all possible past influences of all the amplitudes distributed
throughout space.  Classical trajectory and causal influence are
realized only because the influence of the rest of the paths
sum over to zero due to the rapid phase fluctuations, an obvious characteristic
of the resonant selection of laws.

Other related evidences that physical laws, both classical and quantum,
are globally and resonantly selected include the fact that most laws
are deriveable from the least action or variational
principles\cite{Goldstein1980}.  This otherwise mysterious characteristic
can now be understood in that every physical process simultaneously
samples the environmental/boundary conditions of the entire space
of relevance to filter in the surviving resonant component, which turns out
to be the most efficient (in the least action sense) among
all possible realizations of the process.

In the following subsections we discuss other features of physical
laws which reveal their origin as emergent properties in the relational
dynamics of the universe.

\subsection{Symmetry and Conservation Relations}

The well-known relation between the symmetry of a dynamical system
and the corresponding conservation law that holds for such a system
is the celebrated Noether's theorem\cite{Noether1918,Goldstein1980}.
The existence of Noether's theorem
is another indication that physical laws and the
matter contents have mutual dependence.
This mutual dependence is ingrained in the forms of the relevant dynamical equations
(which is the reason Noether's theorem can be proven by using these
equations) since the laws/equations and the matter distribution
are co-selected out of the the universe resonant cavity.
                                                                  
For the large-scale distribution of matter in the universe, we have
the approximate time invariance (which leads to energy conservation)
and isotropy (which leads to momentum and angular momentum conservation).
However, if we look more closely, both symmetry and conservation
on the large scale are indeed only approximate. 
                                                                     
First of all, the expansion of the universe violates the time invariance
of the matter distribution.  This could have several consequences.
If the values of the fundamental constants are determined by the
characteristics of the universe resonant cavity, depending on the
fashion of this determination the expansion could lead to the variation
of the values of these ``constants'' with time, though the case is not
settled yet as to we have actually observed any such changes\cite{Barrow}.
Another consequence is that over the long time span of the cosmic age,
energy conservation is no longer guaranteed, since the matter distribution
changes with time as a result of the expansion of the universe.
This might be underlying the origin of dark energy and the accelerated
expansion of the universe\cite{Reiss1998,Perlmutter1999}.
                                                             
Secondly, as revealed by the work of Lee and
Yang\cite{Lee1956}, as well as Wu et al.\cite{Wu1957},
in weak interactions parity conservation is violated, i.e., the laws
of physics have a preference for ``handedness''.  If there is
indeed the interrelation between laws and matter distribution,
this handedness in the laws
reveals that large-scale matter distribution in the
universe has a helical component in it.  This would be natural to expect
for a matter distribution originating from the Big Bang or other
initial conditions for the cosmos.
The fact that the parity nonconservation only manifests in weak interactions
perhaps has to do with the fact that weak interaction is the
shortest in range of all the fundamental forces, thus is less sensitive
to the influence of matter distributions in its immediate environment, so other infinite-range
influences which reflect the asymmetries of the universe are more
obviously manifested.  Furthermore, the proven conservation of 
CPT\cite{Pauli1955}, and the demonstration of CP violation 
in certain sub-classes of weak interactions\cite{CP}
show that time-reversal symmetry is violated in these interactions as
well.  Another example of the approximate symmetry and conservation
relation can be found in \cite{Sakurai1967}. Therefore, 
even at the microscopic level we have the evidence of time's arrow manifest,
which is another indication that irreversible evolution and phase transitions
have played a role in the formation and selection of microscopic laws.

These all lead us to conclude that the fundamental resonances
of the universe, when regarded as ``dissipative structures'' in the
universe cavity, are meta-stable dynamical equilibrium structures
in slow secular evolution over the long time span of the history of
the universe.
                                                                     
\subsection{Structures of S-Matrix Theory and Quantum Field Theories}

S-matrix theory was invented to circumvent certain problems
of quantum field theories\cite{Chew1961}.
The motivation for this approach comes from the observation that in many scattering
experiments light quanta come in and go off as approximate plane waves.
In S-matrix theory the dynamics was not specified by a detailed
model of interactions in spacetime, but was determined by the
singularity structure of the scattering matrix, subject to
the requirement of maximal analyticity.  The success of the
S-matrix approach is likely to be due to the fact that the
underlying physics obeys global, modal characteristics.
The results of the calculations are expressed through particles
on the mass shell, which are equated in our ontology to the outcome
of phase transitions.  Other features of the applications of
S-matrix, such as boot-strapping and ``nuclear democracy'' in
hadron theory also reveal distinctive modal characteristics\cite{Cao1998}.
                                                          
Quantum field theories describe local interactions between
particles and fields.  However, certain global elements are
implicit in their formulation for processes such as scattering,
the emission and absorption of photons by atoms, etc.  Some of the common practices in 
quantum field calculations, such as Feynman's diagrammatic approach
are integral representations of the entire phase transition process,
described in terms of input and output states only, and omitting 
(or ``integrating out'') any detailed description of the ``on-location''
behavior of the interaction and particle creation/annihilation.
                                                             
The need for renormalization (or for the manual incorporation
of the experimentally observed values of parameters into perturbative
quantum field theories to cancel certain infinities) is a
reflection of the incompleteness and non-self-consistency status of
the quantum field theories.  The cause of this is partly because
these theories are local (and the renormalization allows the incorporation of
certain environmental screening effects), partly because
they attempt to synthesize quantum theory with special relativity
(the spontaneous quantum phase transitions implicit
in the phenomena that quantum field theories attempt to describe,
such as the creation and annihilation of particles,
necessarily violate Lorentz invariance required by special relativity),
and also partly because global quantum interactions cannot be
truly described as unitary processes\cite{Hagg1,Hagg2}.
                                                             
Gauge field programme emerges within the framework of quantum
field program and incorporates certain global features of field
theories.  Fundamental interactions are characterized by
gauge potentials and the phase of a wave function is
a variable in these theories.  In this formulation local gauge invariance
is used as a criterion to derive the forms of dynamical interactions, and just as with
the variational procedure for the derivation of physical laws,
the success of the gauge invariance procedure reflects the resonant origin
of the relevant laws which imprints on these laws various internal (group transformational)
symmetries.  Spontaneous symmetry breaking was developed as a mechanism to
preserve gauge invariance when dealing with massive gauge quanta,
which, as another example of the phase transition process is
inherently non-unitary.

The idea that the high-energy
effects in gauge theories can be calculated without taking the cutoff
in the normal renormalization procedure to infinity leads to the development
of the Effective Field Theories (EFTs) in the 1970s.  The cutoffs in these
EFTs\cite{Weinberg1980} serve as
boundaries separating energy regions which are separately
describable by different sets of parameters
with different symmetries: for example, QED can be considered an EFT
of the Electroweak theory.  This practice is also related to the so-called
renormalization-group approach which explicitly deals with
the scale-dependence of certain fundamental physical interactions
by introducing the concept of the running coupling strength.
As commented by Cao\cite{Cao1999}, most of the physical theories we know
are scale-independent within their respective range of validity, and the scale-dependence
of parameters in field theories indicates the screening effect of the
environment (which, in our scenario, is the averaged boundary condition
affecting the particular interactions) as first suggested by Dirac, and also indicates the
smoothness and homogeneity of the variation of these interactions with scale.
                                                                     
\subsection{Further Comments on Superluminal Signal Propagation,
Preferred Frame, and Generalized Mach's Principle}

In classical physics, the influences of the nonclassical paths,
which involve superluminal signal propagation, cancel each other
out\cite{Feynman1965}, so the superluminal effect is not apparent
(although as we will show later in the example of the double-slit fringe
formation, if we assign a realistic meaning to the different interfering
paths in classical optics, then the speed-of-light limit is violated there as well).
For quantum physics, in {\em both} the measurement and the
non-measurement type of problems, the effect of space-time paths requiring
superluminal signal propagation can no longer be ignored.
The quantum measurement processes, especially the
Einstein-Podolsky-Rosen type\cite{Bell1,Bell2} or delayed-choice type\cite{Scully1997}
experiments, show both the global extent of the wavefunction
and the superluminal nature of the wavefunction collapse.
That the element of superluminal signal propagation is needed also
during the unitary evolution stage
of a quantum state is less obvious, but is reflected nonetheless in
the quantum mechanical rules of state evolution (equations \ref{eq:1}, \ref{eq:2}), 
in Feynman's path integral formulation of quantum
mechanics (equation \ref{eq:fm}, which contains the integration over
the classical paths at subluminal speed and over nonclassical paths
at superluminal speed), as well as in Feynman diagrams for most of the virtual paths.
Another indication on the instantaneous nature of certain interactions is given
in \cite{Jackson}, where it is shown that the electric field of a charge moving at constant velocity,
viewed by an observer at any distance, is directed along the direction to the current
position of the charge, instead of along the direction to the charge's position at
a past time when the field was emitted. Similar phenomena exist in gravitational
physics as well.

For states which are in the process of going through a phase transition,
such as many quantum tunneling processes and certain optical
evanescent-wave propagation, the speed-of-light limit is often found to be
violated\cite{Steinberg1993,Enders1993}. 
The fact that the superluminal phenomena appears in tunneling type
phenomena is not surprising: here we are in the regime of mesoscopic
coherent resonance formation, therefore 
we have the opportunity to have
a peek at the operation of a kind of zeroth hierarchy process.  
Tunneling and evanescent-wave propagation also involve the exponential decay
regions of the wavefunction distribution in addition to the traveling-wave regions, and
thus the role of phase velocity (which can be orders of magnitude larger than
group velocity even in classical optics analysis) becomes more important.
Another example of faster-than-light
propagation is that of Bell's inequality test\cite{Bell1,Bell2}, where the separation
distance is macroscopic and the superluminal signal propagation feature is revealed.
The predomiment presence of the superluminal signal propagation during quantum
mechanical wavefunction collapse, due to their happening in the microscopic
domain, is not always apparent to the human observer without a careful
contemplation of the physical essence of the process.  We could have mistaken
many of these measurement processes as involving only the local
apparatuses without realizing the role played by the universal
environment in any type of quantum phase transition process.
                                                                    
The existence of the superluminal communication in the zeroth hierarchy
could not be used by humanbeings to send information with, because we
have no control over the zeroth hierarchy dynamics.  To us the zeroth
hierarchy processes will always possess a random element such as exists in
the wavefunction collapse during quantum measurement.

The need for a universal reference frame in quantum mechanical formulation
has become more apparent in recent years\cite{DT1,DT2,B1,Eb1}.
Our current paradigm admits such a preferred universal reference
frame which is simultaneously consistent with the relational nature 
of the origin of physical laws.  With the existence of such an absolute frame,
spacetime events (described by wavefunctions) are uniquely anchored in this frame and 
are not affected by observers' relative motion.  The wavefunctions transform like scalors
and their values are well-defined in whichever frame one choses to observe in,
and the values of these spacetime wavefunctions can always be referred back to
the absolute frame for their causal connection in the event of ambiguity.
The situation then becomes similar to that in observational cosmology, where
even though we presently observe events which happended in the distant past,
it does not mean that these events become truly ``simultaneous'' with events
that are happening around us which we also observe at the present moment, because
we have in some sense an absolute frame established by the isotropic expansion
of the universe so we can register the spacetime events uniquely and can always
place the correct causal context on them as long as we know the relative
spatial locations of them in the absolute frame.

The absolute reference frame also provides an absolute time reference $t_0$.
Space and time thus have quite different characterstics in our current paradigm,
a situation in sharp contrast to the view offered by the relativity theories.  
At every instant $t_0$ the physical configuration in the absolute space
frame is unique (though in general the wavefunction is not at rest with respect
to the absolute space frame), so which detector first encounters the 
wavefunction wavefront in the quantum measurement problem is also unambiguous.  
Each instant of local time corresponds to a unique $t_0$ in the absolute
frame, so the collapse happens based on the boundary condition in the
entire universe at that instant, in a superluminal fashion.
There is thus no longer an issue of ``relativity of simultaneity'' (Appendix A).

The quantities that transform as 4-vectors in the covariant formulation include
interaction quantities which affect the evolution
of the wavefunction as averaged potentials.  The wavefunction collapse phase transition,
however, happens in the zeroth order hierarchy, and is deterministic as far as the boundary
conditions of the universe are deterministic.  Yet the exact result of this
deterministic wavefunction collapse is not predictable because of the
intractability of the many degrees of freedom boundary conditions and the
average nature of the potential. 

Other physical phenomena, apart from the quantum mechanical wavefunction collapse,
which demonstrate the need for an absolute frame include the Aharonov-Bohm 
effect\cite{AB}, which shows that there can be influences to 
the phase of the wavefunction in regions where the nominal field strength
is zero, demonstrating the substantiality of a global potential function
as well as the substantiality of the wavefunction phase which is anchored
in the univeral frame; furthermore, the well-known Josephson effect\cite{JS} in solid
state physics demands a similar substantial and global existence of the quantum mechanical
wavefunction phase.
                     
We have already commented of the fact that GR, as was finally realized
by Einstein, was only partially Machian in nature.  Even though its kinematics
and dynamics are fully relational, there does not exist a prescription for obtaining
mass and inertia from the influence of additional mass in the surroundings.
We now see from our current perspective that such a prescription for calculating the values
of {\em any} dimensional fundamental constants
might never be available to humanbeings because we are not privy to the zeroth hierarchy
laws which generate fundamental constants.
Mach at the time was not aware of the quantized nature of the fundamental interactions which in turn is
a manifestation of the global and instantaneous influence of all the matter and
energy in the universe. In light of our current understandings, 
the fundamental dimensional constants such as c, G, h, e
and $m_e$ could no longer come from a ``theory of everything'' based on 
first hierarchy dynamics.
These constants are the product of zeroth hierarchy dynamics, of
which our current pursuit of science little no glimpse into.
Combinations of these constants, however, can be determined from theoreis of first-hierarchy
physical processes and compared with values obtained from experiments\cite{LP}.

\section{Examples from Quantum Mechanical Experiments}

We now look at several examples of quantum experiments to
illustrate that the new view on the nature of quantum processes can
naturally explain the known peculiarities in the quantum world.

\subsection{Double-Slit Interference, Delayed Choice, and Quantum Eraser}

The double-slit interference experiment is the one of the most-often cited
examples to illustrate the essential features of the quantum measurement process.
In the classical regime, or when the source flux is strong enough that the
arrival of photons is continuous, interference fringes of different orders will
appear on the detection screen behind the double slits.
However, as one lowers the intensity of the source so that on average only one photon at
a time can pass through the slits and reach the detection region, even though
single photons appear to register randomly on the screen during short time span, with long data gathering
time an interference pattern identical to the classical one will be built up.

The question naturally arises as to whether in the single-photon interference regime,
the individual photon went through only one slit, or both slits at the same
time before it reached the detector.  If the photon in question went through only a single slit, 
then how is one to understand the interference phenomenon which
has to be contributed by both slits; or if it went through both slits, 
how did a photon as a particle manage to do this?

A more dramatic realization of the double-slit experiment is Wheeler's Delayed-Choice Experimemt\cite{WL1,WL2},
in which one can make the detection choice of either observing interference fringes (as in the
screen setup), or observing photons from a single path only (if one uses a detector with small
enough field of view so that it collects photons only from the direction of one of the paths),
AFTER the photon has passed through the double slits, so presumably the photon has made the choice of which slit to go
through, if it indeed goes through one slit at a time, or else has gone through both slits at the same time.
So in the conventional Copenhagen interpretation one will have to accept that the act of 
measurement influences the {\em past decision} of the
photon's either going through one slit or the other, or both, which is a clear violation of causality.

Wheeler himself later pushed the paradoxial nature of the violation of causality
to the extreme by imagining the source being a quasar in the distant universe, and the two slits
being realized by two large galaxies serving as gravitational lenses positioned
in the intermediate distance from us {\em en route} of the 
quasar's light propagation. The same chain of arguments of delayed choice observing
then leads to the absurd conclusion that our current choice
of observing a photon from the quasar either as interference from both
galaxies' paths, or from the direction of just one of them, influences the PAST history
of this photon's either taking one or both paths!  This astonishing conclusion has led
Wheeler to invent the phrase of ``genesis by observership'', by which he suggests that our
observations might actually contribute to the creation of the past history of physical reality\cite{TF}.

In our new ontology, however, such counter-intuitive and seemingly absurd conclusions can be totally avoided.
Here we view the detection process as quite a separate physical process
from the propagation process in that it actually re-creates a quantum particle (or else absorbs the
energy equivalent of a quantum particle at an appropriate location) based on the information
propagated to the detector, the detector itself, as well as the rest of the universe, which together form
a resonant cavity.  Likewise, in the emission process a quantum particle is created under the
combined boundary condition of the source and the rest of the universe, including the rest of the
experimental setup.

Therefore what arrives at the detector is the ``influence field'' from the source, which forms only part of the
boundary condition for the detection of the photon.  The detected photon is re-created
partly from the energy in the incoming field, partly from the rest of the universe. In a case like
the double-slit interference where there are two branches of the wavefunction, after the phase transition
occurs at one location the rest of the wavefuction can be considered either as given back to the universe,
or else one can say that the other part of the spread-out wavefunction goes to the 
whole detected photon at the image location: it all amounts to the same
thing in the end, since the modal content, or the quantum, has to be whole, and the measurement phase
transition is instantaneous or superluminal so it can draw energy from where appropriate.
The creation of a new quantum particle has to involve the adjustment of the environment because
a quantum particle is only defined with respect to the enviornment (i.e., a mode cannot be separated
from the substratum it emerges in).  Even a free electron propagating in vacuum will excite Dirac-type
polarization wake.  The ``detection as new creation of particles'' idea can also explain quite naturally
the controversial Afshar experiments\cite{AFS}.

We emphasize that in both the traditional double-slit as well as the double-slit delayed choice experiments
mentioned above, as long as both slits are open, the photon influence field passes through both
slits {\em en route} to the detector.  When the choice of detection mode is made, in one case it
appears  ``as if'' the photon has gone through a single slit only.  However, what is been instantaneously created
is not the {\em past history} of the photon trajectory, but rather the {\em current collapsed modal configuration}
of the detection phase transition.

In one of the delay-choice quantum eraser experiments\cite{Kim}, which realized the original proposal
of \cite{SD}, entangled photon pairs were used to obtain the ``which path'' information as well as
the interference fringes, with the interference fringes recorded before the ``which path''
information was recorded.  The results of these experiments show that
if the correlation is selected such that the which-path information is
available, then the fringes disappear, whereas if such information is ``erased'' 
by selected correlation among the recorded data, then the fringes will reappear.  
This case differs from the straightforward delayed-choice
experiment in that the true choice of interference-or-not of the photon influence field (which is 
a state that must be satisfied by both of the entangled twins) is made randomly at the slit.
In the case of the ``erasing'' output, interference has occured and the field 
contains the coherence needed for potential interference,
it is the particular choice of the mode of correlation by summing together the two sets of fringes with one-half
wavelength offset that has erased the interference.  However, when a photon field indeed
has passed through only a single slit due to random decoherence process at the slit, then there is an 
effective wavefunction reduction before the detection process has taken place.
So unlike the classical delayed-choice implementation, here it is not the detector field of view that leads to the
effective selection of photon field from one slit or both; in this case it happens rather by
either random decoherence at the slit, or else by the particular choice of correlation.
A clear ontological reality can be assigned to any of the scenarios in this experiment.

One other feature of the {\em conventional}
double-slit experiment which can be explained quite well in the new ontology
is the higher-order interference peaks in the interference pattern on the screen.  Whereas for the
zeroth order peak the propagation time from both slits are the same, for the higher-order peaks
the propagation time (or the total path-length) corresponding to each slit path is different.  This leads
to the question of how the photon can travel the two different paths with different amounts of propagation
time and still manage to interfere in the end on the screen and become a whole photon during the
detection process.  In the new ontology, what is being propagated is the influence field, and this
influence field, as the branch-fields in Feynman's path-integral
approach to quantum and classical mechanics, has the different branches propagating in speeds
that are needed to reach a test location simultaneously (many paths thus with superluminal speeds), and is not
constrained by the conventional propagation speed limitations. Natually the coherence length of the photon
field has to be adequate in order to display higher-order fringes.

\subsection{Independent Laser Source Interference Experiments}

In \cite{PM1, PM2} two independent laser sources, rather than a double-slit screen and a single
coherent source, were used to generate interference fringes over a narrow
band.  In subsets of these tests the fringes were observed when the photon flux was so low that only one photon at a
time reached the detector on average before the arrival of the next one. Even 
in this case, however, one cannot infer that the single photon came only from one of the two laser
sources, because turning one of the sources off made the interference pattern disappear.

In the new ontology, all features of the independent light source interference experiment
can be easily explained.  The fact that interference fringes can be produced by the independent
light sources is because the radiated photon energy from either source joins the univesal photon
background as soon as an emission is made, and each one also establishes a unique phase relationship 
within the universal reference frame, thus even though the relative fringe phase is observed to fluctuate
in these experiments because of the fluctuations in the phase of the photon packet
for each emission, the two beams nontheless possess fixed phase relationships during short
measurement intervals.  In \cite{PM2}, the frequency of one of the emitting sources could even
be made slightly different from that of the other source, as long as the source is simultaneously
set in motion so that the Doppler-shifted frequency for the second source 
as viewed by the detector agrees with the
frequency of the stationary source to within the coherent time requirement:
interference fringes are again observed in this latter case.

In \cite{PM1} and \cite{PM2} the authors had concluded that the interference due to independent sources is caused
by the action of the detection.  Our view in this regard differs from these authors', and
this difference can be tested with a ``delayed-choice'' realization of the independent
source interference experiment in the future.  The success of the delayed-choice version of the independent
light source interference experiment, like that of the usual implementation of delayed-choice experiement
with the single source shining through double slits, will show that a time's arrow of energy flow 
from the two sources is an important factor contributing to the final detection result, and the interference
pattern was created in space independent of whether the detector was there or not.  The detection mode of
the detector at the time of the detection did not recreate the {\em past} history of emission,
though if the detector was present at the time of the emission it might have slightly pulled the laser
emission frequency (i.e., the detector
would have been part of the boundary condition of the emission process).
What the detector creates at the time of detection
is only the {\em present} state of the detected photon.

Furthermore, the fact that turning one light source off eliminates the interference completely
shows that both sources' fluxes must be present to enable interference, even if one photon
at a time is detected: this point poses no difficulty in the new ontology because the radiation flux sets
up only an influence field, and this field could be of such a low intensity that only the combination
of the beams from both sources together leads to the arrival of certain number of equivalent photons 
at the detector per unit time.  The detected photon is then created through the combination and interference
of the incoming influence field beams, as well as the additional boundary condition provided
by the detector and the rest of the universe.

Another possible realization of this test which could potentially allow us to distinguish
the current ontology with that of the orthodox Copenhagen interpretation is that if the
emission of photon energy from either source is pulsed with a random trigger, according to the current
ontology there should be no interference observed at the detector.
This will show that it is not merely the lack of {\em knowledge}
of which path the photon takes that will influence the outcome of whether we 
observe the interference or not: there needs to be actual {\em energy} passing 
through both paths for interference to remain!

\subsection{Tests of Bell's Inequality}

The successes (barring the existence of loopholes) of the experimental tests of the violation of Bell's 
inequality\cite{Bell1, Bell2} and the revealation of the ``spooky action at a distance'' nature of 
the quantum mechanical interactions by such experiments
are consistent with the ontology of quantum measurement process proposed in this paper, which 
dictates that the wavefunction collapse process is a discontinuous and instantaneous modification 
of the entire global distribution of the wavefunction.  This process is non-unitary
(since it involves the exchanges with the rest of the universe), is superluminal (since it is an instantaneous
global modification of the wavefunction shape), and it happens in a preferred reference frame, i.e., that defined
by the entire matter and energy distribution in the universe.  These characteristics of the wavefunction collapse
are consistent with the recent findings on the inherent requirement for a typical quantum mechanical measurement
process\cite{Bub}.

Besides the polarization-correlation type of Bell's inequality tests\cite{Bell1, Bell2},
energy-time type of Bell's inequalities have also been derived and tested\cite{Franson1, Franson2}.
In these so-called Franson-interferometer type of tests, two correlated photons are generated from
cascade emissions among three successive levels in an atomic system, with the top level having
a much longer lifetime than the middle level (i.e. $\tau_2 \ll \tau_1$).  These
two correlated photons are subsequently directed into two separate (but otherwise
identical) Michelson interferometers, with each interferometer having both a long arm and a short arm.
When the difference in transit time $\Delta T$ between the short and the long arms is chosen to be intermediate
between the two level lifetimes (i.e. $\tau_2 \ll \Delta T \ll \tau_1$, and also when additional
phase shifts are introduced into the long arms of the two interferometers, the coincidence counts at the
two detectors after the two Michelson interferometers were found to display coherent sinusoidal
dependence on the amounts of the phase shifts and also on the product of the total energy separation
of the top and bottom level of the atomic system $\Delta E$ and the Michelson interferometer
transit time difference between the two arms $\Delta T$, a conclusion which contradicts the
predictions of local hidden variable theories. This unique quantum
mechanical behavior of the result of coincidence between two correlated photons was attributed
to the uncertainty in the time of the emission of the photon pair (i.e. due to the large
value of $\tau_1$ in comparison with the imbalance of the arm lengths $\Delta t$).  
In our ontology, the uncertainty in the emission time of the photon pair
translates to the narrow-band nature of the photon influence field, which makes it possible for
the long and the short arms of the two Michelson interferometers to be traversed simultaneously
by the photon fields.  Furthermore, the two photon influence fields going towards the two
Michelson interferometers remain a single entangled space-time mode due to the shortness of $\tau_2$, and thus global
correlations are maintained and coherence behavior in the coincidence counts were found both
theoretically\cite{Franson1} and experimentally\cite{Franson2}.

Equality-type correlation relations have also been derived\cite{GHZ1, GHZ2, GHZ3} for quantum experiments
which could not hold for local hidden variable theories.  In one possible arrangements of such an
experiment, three correlated photons were sent into three otherwise identical Michelson
interferometers with three phase shifts $\phi_a, \phi_b, \phi_c$, respectively, inserted 
into one of the two arms of each of the three Michelson interferometers.  Detailed derivation\cite{GHZ2}
shows that the probability of triple coincidence depends on the three phase shifts through
\begin{equation}
P = {1 \over 8} [1 \pm \sin (\phi_a + \phi_b + \phi_c)],
\label{eq:P}
\end{equation}
where the plus and minus signs apply to one or the other set of the two sets of three coincidence detectors 
in the 6 output ports of the three Michelson interferometers. 
As the simple reasoning in \cite{CKS} shows, a local realistic theory leads to the prediction
that when all the three phase shifts are changed from zero to $\pi/2$, the three detectors corresponding to the minus
sign in the above equation should give a triple-coincidence reading, whereas quantum mechanics predicts
a coincidenec probability of zero according to (\ref{eq:P}).  In our ontology this result is due to that
in quantum mechanics the
state of the three correlated photons is a distributed spatial-temporal mode, and
one cannot switch the state of one without affecting the other, as is possible in local realistic theories.

\subsection{Cavity QED}

The cavity QED experiments carried out over the past few decades\cite{Rabi1, Rabi2, Rabi3} illustrate 
in the artificially generated environment of a compact configuration many features of the interaction 
of radiation with atoms. 

The quantization of the field in effect is a practice of treating quantum radiation as
collective modes.  In the open environment where the atom can access the fluctuating influence of
the vacuum, the radiation conforms more to the general notion of an irreversible process,
due to the fact that the vacuum field with which the radiating atom is interacting has effectively
infinite degrees of freedom\cite{Scully1997}.  In the specially-prepared environment of a cavity, 
since the available modal content is now finite, the emission and absorption processes can get
entangled with the local environment so a coherent oscillation among the different
atomic states is set up.  In the current ontology such a coherent oscillation can be regarded
as a composite mode under the joint boundary condition of the cavity walls and the atom
in the cavity, as well as the atom's radiation field.  As long as the internal coherence in
such a composite modal oscillation remains, different apparent collapse and recovery cycles
can be observed\cite{Rabi1, Rabi2}. 

The emission of a photon into a cavity could be made to slowly approach that of emission 
into the vacuum by gradually increasing the bandwidth of the cavity. This shows the 
possibility for smooth transition 
from a unitary evolution to an irreversible process, just like the progression of integer series from
a finite number to infinity.  A finite number and true infinity are clearly different things,
yet between the two there is no sharp boundary, only a progression of gradual changes.

Like we have commented before, in essence even the unitary process contains elements of irreversibility
in the realistic physical world which is an ever evolving, open and non-equilibrium system.
True irreversibility in the quantum world, as in the classical, appears to be related to the
open and non-equilibrium secular evolution nature of the univeral environment as a whole,
otherwise we will just be dealing with the intractability of a complex system possessing
nearly infinite degrees of freedom, yet is fundamentally reversible.
The local coherent oscillation in this sense is only an approximate perpectual oscillation
(as is true for all non-equilibrium, metastable ``dissipative structures'')
which will eventually decay due to the interaction with the evolving universal environment.

\subsection{Mesoscopic Interference}

The universal resonance picture of quantum phenomena we proposed can naturally explain
the quantum-to-classical transition. The necessary stability of classical objects is provided by the 
quantum-measurement-type phase transitions which form composite modes, 
thus this mechanism provides the nonlinearity
needed to interrupt the reversible and unitary evolution of the Schrodinger type.

In some of the mesoscopic interference experiments reported so far\cite{meso1}, partial collapse of the wavefunction
and a mild degree of irreversibility in the sense of infrared photon emission had been observed, yet the main
modal content and the ability for interference of the complex molecule used in the interference experiment
survived, which demonstrate the robustness
of these composite modes and the flexibility of their structures to small perturbative changes in the environment.
In the case of infrared photon emission, the coherent composite mode co-emits this photon (in some sense from both branches
of the interference paths in the double-slit experiment), and thus the coherent interference phenomenon
is not destroyed.

We comment here that the success of the mesoscopic interference experiment by no means supports the notion
of the traditional measurement theory of von Neumann type\cite{VonNeumann1932} which claims that the incoming physical states
excite in the measurement instrument a macroscopic superposition state of $(spin\_up)*(pointer\_up) + 
(spin\_down)*(pointer\_down)$ type, or alternatively known as the Schrodinger-Cat state.  A mesoscopic coherent state was resonantly
selected by the complicated natural environment, and its structural stability is also a result of this ``natural
selection'' process.  The coherent superposition state between an incoming pure quantum state and
a classical instrument is not something that can be established by, say, a single hit of a single
incoming photon (which is hardly a ``natural selection'' process).  
Thus far, no experimental evidence of such a macroscopic superposition state 
has been found, neither are there any arguments in the fundamental quantum theory that
support the existence of such states.

Finally, we comment that in regard to the question of what type of mesoscopic object is intrinsically capable
of interference\cite{Leggett}, one criterion could be that if such an object is left in free space, it should
have a tendency to ``diffuse'' (in the Schrodinger's free-electron wavefunction sense)
into its momentum eigenstate, i.e., that of plane wave state.  Such a diffusion would allow the
wavefront splitting needed for the double-slit type of interference experiments.

\subsection{Minimum Disturbance Quantum Measurements. Interaction-Free Measurements}

There are several classes of quantum measurement processes that introduce minimal disturbance to the
system being measured.  These include the quantum non-demolition experiments\cite{QND1,QND2},
quantum Zeno and anti-Zeno effects\cite{QZ1,QZ2,QZ3}, and interaction-free measurements\cite{IF1,IF2,IF3}.

A common feature of these experiments is that the interaction with the state wavefunction to be measured is in the
form of gentle modification of the boundary condition for the relevant quantum mode.  There is not a total wavefunction
collapse, yet the reduction (or renormalization)
of the wavefunction has nontheless happened so there is in fact interaction
between the wavefunction and the environment. 
Even in the so-called interaction-free measurement an interaction with the boundary condition has
nonetheless occured.  The apparent interaction-free modification of the wavefunction is due to the
artificial feature of the unitary formalism, which disregards the back-reaction of the wavefunction
evolution on the forcing potential as well as on the ``rigid'' type boundary conditions.

In terms of the influence field picture we introduced in this paper, for the particular type of interaction-free
measurements which involved the placement of a detonable bomb in one path of the Michelson 
interferometer\cite{IF2},
if the equivalant of a single photon's energy propagates through the entire experimental setup at one time
until it reaches the detector, then equivalently only half of the photon's energy reaches into the path
containing the bomb, which should lead to a probability of 1/2 of its being set off (because otherwise
it would go off every time and there is no sense of discussing the results of the experiment any further),
and at those times when it did not go off, the detection process at the other branch of the Michelson
interferometer should ``absorb'' the 1/2 portion of the photon energy from this other path superluminally
as in other types of quantum mechanical wavefunction collapse processes.

\subsection{Collective Effects in Solids and Liquids}

In condensed matter physics the modal formation feature of the second hierarchy processes is
very prominent. Quasiparticles such as phonons, polarons, plasmons, etc. are all instances of
collective excitations under different boundary conditions\cite{Kittel}.
Collective phenomena such as superconductivity and superfluidity are instances of collective
quantum phenomena on mesoscopic scales.

In the Josephson effect\cite{JS} the pairs made up of the supercurrent tunnel just like a single particle because
it is a composite mode.  Tunneling has the easiest interpretation in a modal picture, and the formation
of these modes could ``borrow'' energy from the rest of the universe during transient\cite{CKS}.
The Josephson effect also demonstrates the reality of the phase of the quantum mechanical wavefunction.

Solid state physics has close-in boundary conditions and local ``resonant cavities'' to compete with
the universal one.  The resulting compromise is the ``quasi-particles" which have both
local and global features, such as in fractional quantum Hall effect where electrons
in strong magnetic field acting together and form ``dressed particles''
which appear to possess fractional electron charge\cite{Laughlin99}.


\section{Summary and Conclusions}

Here is a summary of the major implications of this work:

\begin{itemize}
\item The universe we inhabit is finite, with laws that are relational and emergent, and are generated
out of the mutual interactions of matter.
\item The generation of laws is through a hierarchical process, and the process is relational and
emergent throughout the various hierarchies.  
Humanbeings are not privy to the laws operating in the zeroth hierarchy, because these are
instantaneous and global, therefore the actual process of the formation of laws is
not observable apart from its consequences to the first order hierarchy laws and phenomena.
The superluminal communication in the zeroth hierarchy cannot be tapped by humanbeings for
information transmission.  The fundamental physical laws we have obtained belong to the first
hierarchy, so are the transformation relations of laws such as Lorentz transform and general
covariance, which are all emergent properties.
\item The absolute reference frame established by the matter and energy content of the universe is what the
quantum mechanical wavefunction collapses with respect to.  It also anchors the absolute phase of 
the wavefunction in the universe in relation to one another in a global sense.
\item There is strict temporal causality in the absolute frame, but the causal influences are spatially distributed
and self-referential, rather than linear and sequential between cause and effect.
\item The substratum of the quantum world is continuous, and the discrete resonances are due to the
finite nature of the universe resonant cavity and the non-equilibrium phase transitions
happening in it.
\item There is intimate connection between the covariant unitary evolution and the 
superluminal and irreversible wavefunction collapse, in that the potential function in the
Hamiltonian used to drive the unitary evolution reflects the averaged influence from the
results of large numbers of wavefunction-collapse phase
transitions in the environment.  Furthermore, the evolution of the wavefunction has subtle
back-reactions on the spectrum of modes in the universe cavity (i.e., on the environment of
the relevant wavefunction), and in the rigorous sense the
entire universe should be considered co-evolving with continuous and wide-spread wavefunction
collapses. The unitary evolution phase is an idealization and is an emergent process.
\item There does not exist a so-called ``wavefunction of the universe'' because to have a wavefunction
one needs to know the average driving potential to which the back-reaction due to the evolution
of the wavefunction can be ignored. When the universe is considered as a whole there
is only a network of mutually-influencing components and a multitude of phase transitions
in the zeroth hierarchy, and one cannot isolate an average potential in this case which experiences
no back-reactions.  It is thus best to consider the co-existence of 
many different wavefunctions in different hierarchies and different
regions of the physical world, and the environmental influence to the local physics can be
approximated by an average driving potential for the local wavefunction which can itself be considered
as well-isolated apart from the driving influence of the outside potential.
\item In this scenario there exists a ``fundamental forbidden rule'' which prevents any human
constructed ``theories of everything'' to predict individually the values of the dimensional
fundamental constants from first principles, though the {\em relations} among different groups
of fundamental constants can be obtained through theoretical models of the first and second hierarchy
collective effects which incorporate additional experimental input.
\item In the ultimate sense, the universe is an open and evolving system, and the structures
we observe are quasi-steady states existing in dynamical equilibrium.
All the laws and fundamental properties in the universe co-evolve with the matter distribution
in the long run, as befitting to all such nonequilibrium systems.
\end{itemize}

This broad systhesis encompasses all the pheonomena we know in both the classical
and the quantum world, and also accounts for the structures of the successful
dynamical theories we have obtained so far.  The self-consistent account of the diverse range
of physical phenomena and theoretical structures gives this proposal its initial appeal,
and more detailed inferences will be worked out as our understanding
progresses.

\appendix
\section{Superluminal Time Synchronization, Coordinate Transformation 
and Causality in a Hierarchically-Ordered Universe}

We present a detailed example here to illustrate features of the coordinate transformation and 
time synchronization in the different hierarchies of the universe
to demonstrate that temporal causality is well
preserved in this framework. 

Assume two experimenters, Alice and Bob, who are at rest in coordinate system 1 with positions
$x_{A1}$ and $x_{B1}$, respectively.  Without losing generality, we assume $x_{A1} < x_{B1}$.
Now Alice performs a Bell-type measurement at $t_{A1}$ (say, by flipping the polarizer encountering
one of the twin-photons from parametric down-conversion), which impacts Bob's entangled twin
at $t_{B1}$, we obviously have $t_{A1} < t_{B1}$ even though the difference in time might be
imperceptible.

Assume now that another coordinate system 2 is moving with velocity $v \hat{x}$ with respect to
coordinate system 1.  The standard Lorentz transform relation gives
\begin{equation}
t_{B2} - t_{A2} = \gamma [ (t_{B1} - t_{A1}) - (v/c^2) (x_{B1}- x_{A1})],
\end{equation}
where
\begin{equation}
\gamma={{1} \over {[1-(v/c)^2]^{1/2}}}.
\end{equation}

It can be seen from the above experssion that depending on the sign and magnitude of $v$,
$t_{B2} - t_{A2}$ could be either of the same sign or different sign as $t_{B1} - t_{A1}$,
i.e., the causal sequence observed by an observer in coordinate system 2 could be the opposite
of that occurred in coordinate system 1.  It is easily shown that the correct causality sequence
can be preserved if $v$ as well as the signal propagation speed in coordinate system 1
between $x_{A1}$ and $x_{B1}$ are both not exceeding $c$, the speed of light in vacuum.
For then
\begin{equation}
t_{B2} - t_{A2} = \gamma (t_{B1} - t_{A1})[1 - (v v_1/c^2)]
,
\end{equation}
where $v_1$ is the signal propagation velocity (in this case positive because both
$t_{B1} - t_{A1}$ and $x_{B1}- x_{A1}$ are greater than zero)
in coordinate system 1 between the events at $x_{B1}$ and $x_{A1}$.

However, in Bell's inequality type of tests, we do have superluminal signal propagation
between $x_{B1}$ and $x_{A1}$, so for an observer in coordinate system 2, causality (or at least
the {\em apparent} causal sequence of events) are violated. This issue has been realized
by many previous investigators\cite{DT1, DT2}.

With the existence of an absolute frame, it was thought that the ills relating to the causal
loops produced by the standard formulation of quantum mechanics can be cured\cite{DT1,DT2}.  An absolute
frame does establish an absolute order of the occuring of events, if these events are occuring
with respect to the absolute frame.  But what about a causal sequence occuring in, say frame 1,
whereas our absolute frame is moving with $v \hat{x}$ with respect to frame 1 (i.e., our absolute frame
becomes the frame 2 in the above example)?  Then the observer
in our absolute frame with have a distorted view of the causal sequence of the Bell measurement
we have described above.

Therefore, to truly restore the correct causal sequence in the absolute frame formulation, in addition
to having such a universal frame to compare things with, we also need the additional feature
of our zeroth hierarchy process, i.e., we need to have effectively a signal propagation speed
$c=\infty$ in the absolute frame to accomodate the causality issues raised by the
zeroth hierarchy processes (of which quantum wavefunction
collapse and Bell-type measurements are all examples).  With an infinite signal propagation
speed, we have in the above expressions $\gamma=1$, and $t_{B2} - t_{A2} = t_{B1} - t_{A1}$.
Which is to say, the time interval measured by the local observer is exactly the same as the
time interval measured in the absolute frame, even though the absolute value of each instant of
time may be tagged differently in these two frames.  We have therefore gotten rid of the causality
confusion once and forever if we have an absolute frame with instantaneous and
universal clock synchronization!
Similarly, with the effective ``instant messaging'',
the length intervals measured in the absolute frame will be equal to that measured
in the local frame, no matter the motional state of the local frame with respect to the absolute frame.
These results
reflect the fact that in general, the Lorentz transformation relations degenerate into
the Galilean transformation ones when the speed of maximum signal propagation is infinite.

What about the transformation relations between the different inertial frames, apart from the
absolute frame?  The standard Lorentz transformation relations should still hold, and it will
unavoidably occur that some spacetime event sequences are thus scrambled (i.e., space and
time are mixed as viewed by different inertial observers).
In most cases these perceptions just merge into the average influence potential which
enable them to transform covariantly.  In cases where the temporal ordering of
events are crutial, then
just as in the case of observational cosmology, once we have the absolute frame to
anchor ourselves, we can regress the true sequence of events by calibrating our time and position offset
with respect to the absolute frame, as well as the events we observe.  We did not conclude that
just because we observed the microwave background radiation and the life around us all at the same
moment (the present time), that these are events occuring simultaneously, because we have
other indirect knowledge that the microwave background radiation came a long way to reach us!
Similarly, many of the ``relativity of simultaneity'' effects of special relativity are due
to the perception effect caused by the relative motions and separation of the reference
frames and events, which, once we can correctly register with respect to the absolute frame, become
in this frame an orderly procession of events in the absolute sense.

\begin{acknowledgments}
The author acknowledges helpful comments from R. Lucke, L. Sica, and M. Steiner.
This research was supported in part by funding from the Office of Naval Research.
\end{acknowledgments}


\begin{thebibliography}{}
\bibitem{E1} A.~Einstein, \emph{Ann. D. Phys. (4)} {\bf 17}, 891 (Freeman, NY, 1973).
\bibitem{DT1} L.~Hardy, \emph{Phys. Rev. Lett.} {\bf 68}, 2981 (1992).
\bibitem{DT2} I.~Percival, \emph{Phys. Lett. A} {\bf 244}, 495 (1998).
\bibitem{B1} J.S.~Bell, in \emph{Quantum Gravity}, eds. C.J.~Isham, R.~Penrose, and D.W.~Sciama (Oxford Univ. Press, Oxford,
1981), p.611
\bibitem{Eb1} P.H.~Ebergard, {\emph Nuovo Cimento B}, {\bf 46}, 392 (1978)
\bibitem{Bell1} A.~Aspect, J.~Dailbard, and G.~Roger, \emph{Phys. Rev. Lett.}, {\bf 49}, 1804 (1982).
\bibitem{Bell2} W.~Tittel, et al., \emph{Phys. Rev. A}, {\bf 57}, 3229 (1998).
\bibitem{PM1} R.L.~Pfleegor, and L.~Mandel, \emph{Phys. Rev.} {\bf 159}, 1084 (1967).
\bibitem{PM2} R.L.~Pfleegor, and L.~Mandel, \emph{J. Opt. Soc. Am.} {\bf 58}, 946 (1968).
\bibitem{MW} L.~Mandel, and E.~Wolf, \emph{Optical Coherence and Quantum Optics} (Cambridge Univ.
Press, Cambridge, 1995). 
\bibitem{J} V.~Jacques et al., \emph{Science} {\bf 315}, 966 (2007).
\bibitem{WL1} J.A.~Wheeler, in \emph{Quantum Theory and Measurement}, eds. J.A.~Wheeler and W.H.~Zurek,
(Princeton. Univ. Press, Princeton, 1983), 182.
\bibitem{WL2} J.A..~Wheeler, in \emph{Foundational Problems in the Special
Sciences}, R.E.~Butts and J.~Hintikka eds., (Reidel, Dordrecht, 1975), 3.
\bibitem{TF} T.~Folger, \emph{Discover}, {\bf 23}, No. 6, 44 (2002).
\bibitem{SB} S.~Begley, \emph{Newsweek}, Aug. 15 (2007).
\bibitem{AB} Y.~Aharonov, and D.~Bohm, \emph{Phys. Rev.} (Ser. 2) {\bf 115}, 485 (1959).
\bibitem{AC} Y. Aharonov, and A.~Casher, \emph{Phys. Rev. Lett.} {\bf 53}, 319 (1984).
\bibitem{JS} B.D.~Josephson, \emph{Phys. Lett.}, {\bf 1}, 251 (1962).
\bibitem{XZ1} X.~Zhang, \emph{US Naval Research Laboratory Memorandum Report} NRL/MR/7218-05-8883, arXiv/quant-ph/0506100 (2005).
\bibitem{XZ2} X.~Zhang, \emph{Proceedings of Albert Einstein Century International Conference, AIPC} {\bf 861}, 546 (2006).
\bibitem{AE1} A.~Einstein, {\emph Preuss. Akad. Wiss. Berlin, Sitzber.,} 844 (1915).
\bibitem{Mach} E.~Mach, \emph{The Science of Mechanics: A Critical Historical Account of its Development} 
 (1883; 6th ed., Open Court, La Salle IL, 1960).
\bibitem{MTW} C.W.~Misner, K.S.~Thorne,  and J.A.~Wheeler, \emph{Gravitation} (Freeman, NY, 1973).
\bibitem{Wb} S.~Weinberg, \emph{Gravitation and Cosmology} (Wiley, NY, 1972).
\bibitem{Prigogine1977} G.~Nicolis, and I.~Prigogine, \emph{Self-Organization in Nonequilibrium Systems} (Wiley, NY, 1977).
\bibitem{Zhang1998} X.~Zhang, \emph{Astrophysical Journal}, {\bf 499}, 93 (1998).  
\bibitem{Barbour1995} J.~Barbour, and J.H.~Pfister eds., \emph{Mach's Principle: From Newton's Bucket to 
 Quantum Gravity} (Birkhauser, Boston, 1995).
\bibitem{Sachs2003} M.~Sachs, and A.R.~Roy eds., \emph{Mach's Principle and the Origin of Inertia} (Apeiron, Montreal, 2003).
\bibitem{dB} L.~de Broglie, \emph{Non-linear Wave Mechanics}, (Elsevier, Amsteradm, 1960).
\bibitem{BM} D.~Bohm, \emph{Phys. Rev.} {\bf 85}, 166, 180 (1952).
\bibitem{Schrodinger1926} E.~Schrodinger, \emph{Die Naturwissenschaften} {\bf 28}, 664 (1926).
\bibitem{Sakurai1985} J.J.~Sakurai, \emph{Modern Quantum Mechanics} (Addison-Wesley, NY, 1985).
\bibitem{IF1} R.H.~Dicke, \emph{Am. J. Phys.}, {\bf 49}, 925 (1981).
\bibitem{IF2} A.C.~Elitzer, and L.~Vaidman, \emph{Found. Phys.}, {\bf 23}, 987 (1993).
\bibitem{IF3} P.~Kwiat, H.~Weinfurter, T.~Herzog. A.~Zeilinger, and M.~Kasevich, \emph{Proc. Symp. Found. Phys.}, (1994).
\bibitem{VonNeumann1927} J.~von~Neumann, \emph{Gott. Nach.} {\bf 1-57}, pp. 245-272 (1927).
\bibitem{VonNeumann1932} J.~von~Neumann, \emph{Mathematische Grundlagen der
Quantenmechanik} (Spinger, Berlin, 1932).
\bibitem{Cao1998} T.Y.~Cao, \emph{Conceptual Developments of 20th Century Field Theories},
(Cambridge Univ. Press, Cambridge, 1998).
\bibitem{Loudon} R.~Loudon, \emph{The Quantum Theory of Light}, 2nd edition (Oxford Univ. Press, Oxford, 1983).
\bibitem{Sakurai1967} J.J.~Sakurai, \emph{Advanced Quantum Mechanics} (Addison-Wesley, NY, 1967).
\bibitem{Milonni1994} P.W.~Milonni, \emph{The Quantum Vacuum: An Introduction to Quantum
Electrodynamics} (Academic Press, San Diego, 1994).
\bibitem{Dirac1958} P.A.M.~Dirac, \emph{The Principles of Quantum Mechanics}, 4th edition (Oxford Univ. Press, Oxford, 1958).
\bibitem{Brown1954} R.~Hanbury Brown, and R.Q.~Twiss, \emph{Philos. Mag.}, Ser. 7, {\bf 45}, 663 (1954).
\bibitem{Berman1997} P.R.~Berman, ed. \emph{Atom Interferometry} (Academic Press, San Diego, 1997).
\bibitem{Scully1997} M.O.~Scully, and M.S.~Zubairy, \emph{Quantum Optics} (Cambridge Univ. Press, Cambridge, 1997), p.125.
\bibitem{Schrodinger1935} E.~Schrodinger, \emph{Naturwissenschaften}, {\bf 23}, 807, 823, 844 (1935).
\bibitem{LP} R.B.~Laughlin, and D.~Pines,  \emph{Proc. Nat. Aca. Sci.}, {\bf 97}, 28 (2000).
\bibitem{Feynman1965} R.P.~Feynman, and A.R.~Hibbs, \emph{Quantum Mechanics and Path
Integrals} (McGraw-Hill, NY, 1965).
\bibitem{Goldstein1980} H.~Goldstein, \emph{Classical Mechanics}, 2nd ed. (Addison-Wesley, Menlo Park, 1980).
\bibitem{Noether1918} E.~Noether, \emph{Nachr. Akad. Wiss. Goett.}, 235 (1918).
\bibitem{Barrow} J.D.~Barrow, \emph{The Constants of Nature} (Pantheon, NY, 2002).
\bibitem{Reiss1998} A.G.~Reiss et al. \emph{Astronomical Journal} {\bf 116}, 1009 (1998).
\bibitem{Perlmutter1999} S.~Perlmutter et al. \emph{Astrophysical Journal} {\bf 517}, 565 (1999).
\bibitem{Lee1956} T.D.~Lee and C.N.~Yang, \emph{Phys. Rev.} {\bf 104}, 254 (1956).
\bibitem{Wu1957} C.S.~Wu, E.~Ambier, R.W.~Hayward, D.D.~Hoppes, and R.B.~Hudson, \emph{Phys. Rev.} {\bf 105}, 1413 (1957).
\bibitem{Pauli1955} W.~Pauli, in \emph{Niels Bohr and the Development of Physics}, ed. W. Pauli,
\bibitem{CP} L.~Wolfenstein eds., {\emph CP Violation} (North-Holland, Amsterdam, 1989).
L.~Rosenfeld, and V.~Weisskopf (MaGraw-Hill, NY, 1955).
\bibitem{Chew1961} G.F.~Chew, \emph{S-Matrix Theory of Strong Interactions} (Benjamin, NY, 1961).
\bibitem{Hagg1} R.D.~Hagg, 
{\emph Det Kongelige Danske Videnskabernes Selskab, Mathematisk-fysiske Meddeleser}, {\bf 29}, nr. 12, 1
\bibitem{Hagg2} D.~Hall, and A.S.~Wightman, 
{\emph Det Kongelige Danske Videnskabernes Selskab, Mathematisk-fysiske Meddeleser}, {\bf 31}, nr. 5, 1
\bibitem{Weinberg1980} S.~Weinberg, \emph{Phys. Lett.} {\bf 91B}, 51 (1980).
\bibitem{Cao1999} T.~Y. Cao, in T.Y. Cao ed.  \emph{Conceptual Foundations of Quantum Field Theory} 
(Cambridge Univ. Press, Cambridge, 1999).
\bibitem{Steinberg1993} A.M.~Steinberg, P.G.~Kwait, and R.Y.~Chiao, \emph{Phys. Rev. Lett.} {\bf 71}, 708 (1993).
\bibitem{Enders1993} A.~Enders, and G.~Nimtz, \emph{Phys. Rev. E} {\bf 48}, 632 (1993).
\bibitem{AFS} S.S.~Afshar, \emph{Proc. SPIE}, {\bf 5866}, 229 (2005).
\bibitem{Kim} Y.-H.~Kim, R.~Yu, S.P.~Kulik, and Y.H.~Shih, \emph{Phys. Rev. Lett.}, {\bf 84}, 1 (2000).
\bibitem{SD} M.O.~Scully, and K.~Druhl, \emph{Phys. Rev. A}, {\bf 25}, 2208 (1982).
\bibitem{Bub} J.~Bub, arXiv:quantph/0402149 (2004).
\bibitem{Franson1} J.D.~Franson, \emph{Phys. Rev. lett.}, {\bf 62}, 2205 (1989).
\bibitem{Franson2} J.~Brendel, E.~Mohler, and W.~Martienssen, \emph{Eerophys. Lett.}, {\bf 20}, 575 (1992).
\bibitem{GHZ1} D.M.~Greenberger, M.A.~Horne, and A. Zeilinger, in \emph{Bell's Theorem,
Quantum Theory and Conceptions of the Universe}, M. Kafatos ed. (Dordrecht, Kluwer, 1989).
\bibitem{GHZ2} D.M.~Greenberger, M.A.~Horne, A.~Shimony, and A. Zeilinger, in \emph{Am. J. Phys.},
{\bf 58}, 1131 (1990).
\bibitem{GHZ3} D.M.~Greenberger, M.A.~Horne, and A. Zeilinger, in \emph{Phys. Today}, {\bf 46}, 22 (1993).
\bibitem{CKS} R.Y.~Chiao, P.G.~Kwiat, and A.M.~Steinberg, in \emph{Advances in Atomic, Molecular,
and Optical Physics}, {\bf 34}, 35
\bibitem{Rabi1} G.~Rempe, H.~Walther, and N.~Klein, \emph{Phys. Rev. Lett.}, {\bf 58}, 353 (1987).
\bibitem{Rabi2} F.~Bernardot, P.~Nussenzveig, M.~Brune, J.M.~Raimond, and S.~Haroche, \emph{Europhys. Lett.}, {\bf 17}, 33 (1992).
\bibitem{Rabi3} S.~Haroche, and D.~Klepper, \emph{Physics Today}, Jan. (1989), p.24.
\bibitem{meso1} M.~Arndt, O.~Nairz, and A.~Zeilinger, in \emph{Quantum Unspeakables} (Springer, Berlin, 2002), p.333.
\bibitem{Leggett} A.J.~leggett, \emph{J. Phys. Cond. Matt.}, {\bf 14}, R415
\bibitem{QND1} G.~Nogues, A.~Rauschenbeutel, S.~Osnaghi, M.~Brune, J.M.~Raimond, and S.~Haroche, \emph{Nature},
{\bf 400}, 239 (1999).
\bibitem{QND2} C.~Guerlin, et al., \emph{Nature}, {\bf 448}, 23 (2007).
\bibitem{QZ1} B.~Misra, and G.~Sudarshan,\emph{J. Math. Phys.}, {\bf 18}, 756 (1977).
\bibitem{QZ2} W.M..~Itano, D.J.~Heinzen, D.J.~Bollinger, and D.J.~Wineland, \emph{Phys. Rev. A}, {\bf 41}, 2295 (1990).
\bibitem{QZ3} M.C..~Fischer, B.~Gutierrez-Medina, and M.G.~Raizen, \emph{Phys. Rev. Lett.}, {\bf 87}, 040402-1 (2001).
\bibitem{Kittel} C.~Kittel, \emph{Introduction to Solid State Physics}, (Wiley, NY, 1986)
\bibitem{Laughlin99} R.B.~Laughlin, {\emph Rev. Mod. Phys}, {\bf 71}, 863 (1999).
\end{thebibliography}
\end{document}